\documentstyle[prl,aps,epsfig]{revtex}         
\begin{document}
\draft               
\twocolumn[\hsize\textwidth\columnwidth\hsize\csname @twocolumnfalse\endcsname

\title{Spatial solitons in a semiconductor microresonator}
\author{V.B.Taranenko, I.Ganne,$^*$ R.J.Kuszelewicz,$^*$ C.O.Weiss}
\address{Physikalisch-Technische Bundesanstalt 38116 Braunschweig/Germany\\
$^*$Centre National d'Etudes de Telecommunication, Bagneux/France}
\maketitle
\begin{abstract}
We show experimentally the existence of bright and dark spatial
solitons in a passive quantum-well-semiconductor resonator of
large Fresnel number. For the wavelength of observation the
nonlinearity is mixed absorptive/defocusing. Bright solitons
appear more stable than dark ones.
\end{abstract}
\pacs{PACS 42.65.Sf; 42.65.Tg; 42.70.Nq} \vskip1pc ]
The possibility of the existence of spatial solitons in
semiconductor resonators has recently been investigated in some
detail theoretically \cite{tag:1,tag:2} motivated, among others,
by possible usefulness of such structures in new types of optical
information processing. Bright and dark solitons have been
predicted. The majority of the papers treated bright solitons
\cite{tag:1}. We have recently reported on experiments concerning
the switching space-time dynamics of quantum-well semiconductor
resonators \cite{tag:3}. In these experiments we showed the
existence of hexagonal patterns, some "robustness" of small
switched domains and the possibility of switching of individual
elements of ensembles of bright spots; thus giving first evidence
for stable localized structures. In this letter we show directly
the existence of bright and dark solitons. \\

The semiconductor resonator used for the measurements consists of
flat Bragg mirrors of about 99.7 $\%$ reflectivity with 18
GaAs/Ga$_{0.5}$Al$_{0.5}$As quantum wells between them
\cite{tag:4}. The optical resonator length is approximately 3
$\mu$m, the area about 2 cm$^2$. Across this area the resonator
wavelength varies, so that one can work from the dispersive range
(wavelength longer than the gap wavelength) to well within the
absorption band. The resonator was optimized for dispersive
optical bistability, for which reason the absorption of the
semiconductor material is too high for absorptive bistability at
the bandgap wavelength (the most stable solitons are predicted for
absorptive bistability, i.e. for wavelengths near band gap). \\
Our previous experiments in the dispersive bistability region
\cite{tag:3} had shown linear structure formation in the form of
somewhat irregular clusters of bright spots. These are simply the
result of filtering of light scattered in the material, by the
high Fresnel number, high finesse resonator \cite{tag:5}. This
structured background field prohibits the formation of pure
unperturbed, independent solitons. We attempted therefore to work
closer to the band gap wavelength, where the absorption of the
nonlinear quantum-well material is larger and thus the resonator
finesse smaller, in order to avoid the linear structuring of the
field. \\ The experimental set up is conceptually simple: light is
generated by a Ti:Al$_2$0$_3$-laser tunable to desired
wavelengths. It irradiates in a spot about 40 $\mu$m diameter the
semiconductor sample, with intensities of order of kW/cm$^2$ as
required for saturating the semiconductor material. As the
substrate of the semiconductor resonator is GaAs, which is opaque
in the wavelength range in question, the light reflected from the
sample is observed. \\ Observations are done either by a CCD
camera for recording 2D images or by a small detector which can
record the time variation e.g. on a cross-section of the
illuminated area ("streak-camera" - images). In order to avoid as
much as possible thermal nonlinearities, the measurements are done
during a time of a few microseconds. For this purpose the light is
admitted to the sample for about 5 $\mu$s, repeated every ms,
using a mechanical chopper. For recording 2D images with good time
resolution, in front of the CCD camera an electro-optical
modulator is placed as a fast shutter, permitting recording with
exposure times of down to 10 ns. The shutter is triggered after a
variable delay with respect to the beginning of the sample
illumination. Thus the evolution of the reflected light field can
be followed in 2D when varying the trigger delay. \\ Optical
"objects" (e.g. spatial intensity variations) in the light field
move according to the field gradients in phase and intensity
\cite{tag:6}. Thus, as long as there are well-defined gradients in
the field, the time evolution of the 2D field is completely
reproducible in each successive illumination. Consequently,
averaging over several illuminations is possible to increase
signal/noise ratio. The CCD camera in the usual TV-format reads
out a frame every 40 ms i.e. it averages 40 illuminations. For the
finite extinction ratio of the electro-optical modulator (300) we
used a shutter aperture time of 50 ns. The latter can be chosen by
the length of a Blumlein line driving the electro-optical
modulator. \\

Fig. 1 shows the structures observed at $\lambda$ = 860 nm, i.e.
$\approx$ 10 nm in wavelength above the band gap and 5 nm above
the exciton line center. Fig. 1b) shows a bright soliton (dark in
reflection) on an unswitched background, 1c) shows a dark soliton
(bright in reflection) on a switched background. For clarity 1a)
shows a switched area without soliton.
\begin{figure}[htbf]
\epsfxsize=75mm \centerline{\epsfbox{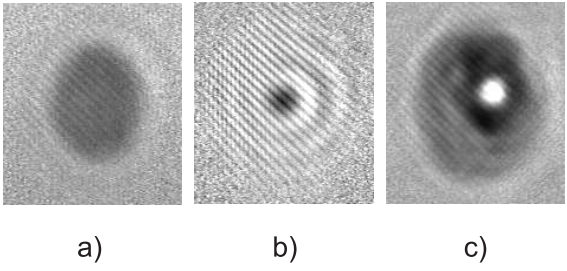}} \vspace{0.5cm}
\caption{Switched structures: reflectivity (reflected
light/incident light) of the sample: a) switched domain (limited
by a contour of Maxwellian intensity) at $\delta \lambda$ = -0.45
nm, power 75 mW; b) bright soliton (dark spot in reflection) at
$\delta \lambda$ = -0.75 nm, power 160 mW; c) dark soliton (bright
spot in reflection) at $\delta \lambda$ = -0.60 nm, power 360
mW;\\ Parameters are: $\lambda$ = 860 nm, size of illuminated area
40 $\mu$m.}
\end{figure}
This switched area is surrounded by a switching front which (when
raising the intensity of the illumination with a Gaussian laser
beam) initially travels outward from the center of the beam until
it stops at an intensity contour given by the Maxwellian intensity
[7].
\begin{figure}[htbf]
\epsfxsize=72mm \centerline{\epsfbox{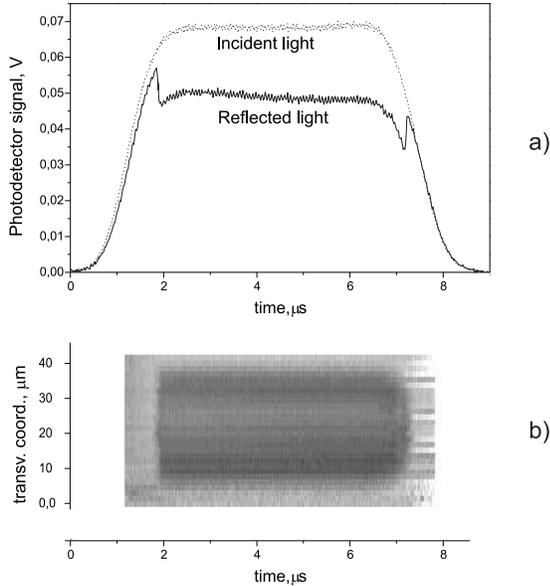}} \vspace{0.5cm}
\caption{ Development of switched domain corresponding to Fig. 1a:
a) incident (dashed) and reflected (solid) light intensity
measured at the center of the Gaussian beam, b) evolution of
switching front (transverse coordinate vs. time).\\ Parameters as
in Fig.1a. }
\end{figure}
The bright soliton exists on a background corresponding to the
lower branch of the (plane-wave) bistability characteristic and
the dark soliton on a background corresponding to the upper
branch. The solitons Fig. 1b),c) develop above the illumination
intensity which switches to the upper branch.\\ Figs 2,3,4 give
"streak-camera" recordings of the formation of the structures Fig.
1. Formation of the switched domain 1a) is shown in Fig. 2. At
time t $\approx$ 2 $\mu$s the resonator switches in the center of
the laser beam and a switching front travels rapidly outward. It
stops, because it reaches the Maxwellian intensity contour and the
domain retains its size until the reduction of illumination
shrinks the domain size and finally switches the resonator back to
the lower branch. \\The development of the bright soliton Fig. 1b)
is given in Fig. 3. The resonator switches at t $\approx$ 1.5
$\mu$s and the switched-up domain develops as in Fig. 2. The
incident light intensity here reaches a higher value than in Fig.
2, which apparently makes switched-up state modulationally
unstable. The consequence is a subsequent shrinking of the
switched-up domain to a small size which is stable. \\
\begin{figure}[htbf]
\epsfxsize=72mm \centerline{\epsfbox{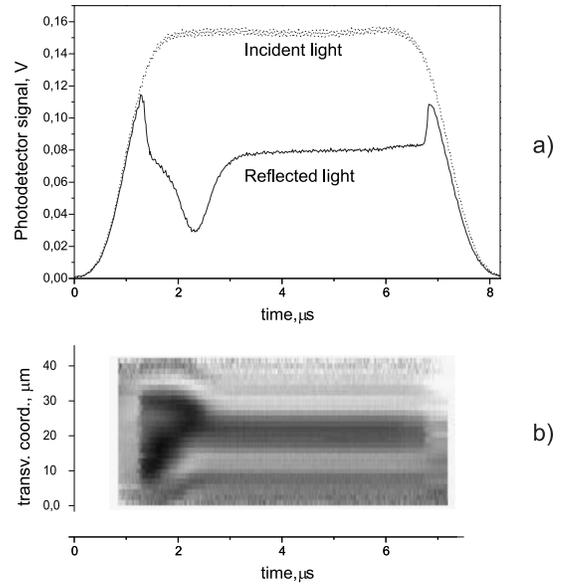}} \vspace{0.5cm}
\caption{ Formation of bright soliton corresponding to Fig. 1b: a)
incident (dashed) and reflected (solid) light intensity measured
at the center of the Gaussian beam, b) evolution of switched
structure (transverse coordinate vs. time). \\ Parameters as in
Fig.1b.}
\end{figure}
We note several features of the stable structure which show its
soliton properties: \\ 1) Stable diameter in time; \\ 2) Size of
10 $\mu$m as expected from model calculations \cite{tag:1,tag:2};
\\ 3) Surrounded by characteristic rings due to the "oscillating
tails" of the switching front \cite{tag:8}; \\ 4) Robustness: From
t = 6.3 $\mu$s the incident light intensity drops , until the
structure switches off at t = 6.8 $\mu$s. In spite of this change
of illumination, the brightness of the structure remains constant.
Such immunity against external parameter variation would seem
characteristic for a nonlinearly stabilized structure; \\ 5) We
note the fast switch-off of the structure at t = 6.8 $\mu$s. The
structure disappears abruptly at a certain intensity, in a manner
suggesting a subcritical process. This allows to conclude that the
nonlinearity is "fast" (electronic). A slow (e.g. thermal)
nonlinearity would not allow such abrupt disappearance of the
structure. \\ Due to the properties 1) to 5) we can identify the
structure with a bright soliton based on a fast nonlinearity. \\
\begin{figure}[htbf]
\epsfxsize=72mm \centerline{\epsfbox{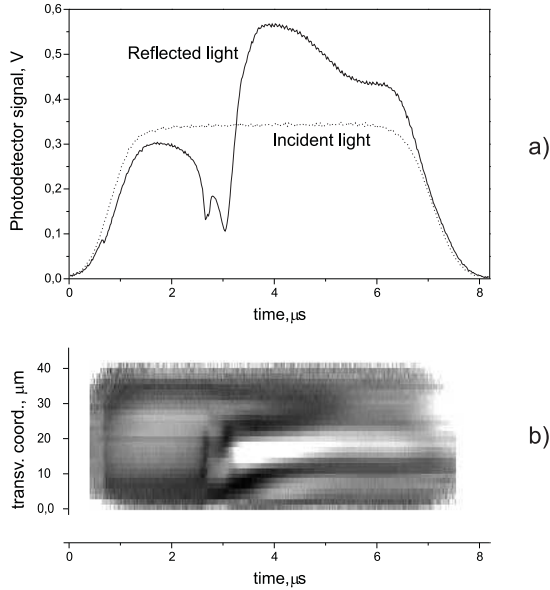}} \vspace{0.5cm}
\caption{ Formation of dark soliton corresponding to Fig. 1c: a)
incident (dashed) and reflected (solid) light intensity measured
at the center of the Gaussian beam, b) evolution of switched
structure (transverse coordinate vs. time). \\ Parameters as in
Fig.1c.}
\end{figure}
The development of the dark soliton structure (Fig. 1c)) is shown
in Fig. 4. The resonator switches at t $\approx$ 0.7 $\mu$s, after
which the illumination intensity is further substantially
increased. Again, after a long transient the bright structure
forms slightly off the beam center (compare Fig. 1c)).
Surprisingly the reflected light intensity of the structure is
almost two times higher than the incident light intensity,
indicating that this structure collects light from its surrounding
(see the dark surrounding of the bright spot in Fig. 1c)), as one
might expect for a nonlinear structure.\\ The reduction of the
brightness of the structure apparent in Fig. 4a),b) in the time
3.5 $\mu$s to 6.5 $\mu$s does not show a damping or disappearance
of the structure but is rather due to a motion of the soliton in a
plane perpendicular to the paper plane. Fig. 5 shows three 2D
snapshots within the time 3.5 $\mu$s to 6.5 $\mu$s clearly showing
this motion. This bright structure on a switched background which
moves like a stable particle and is thus a dark soliton
\cite{tag:2}.\\
\begin{figure}[htbf]
\epsfxsize=70mm \centerline{\epsfbox{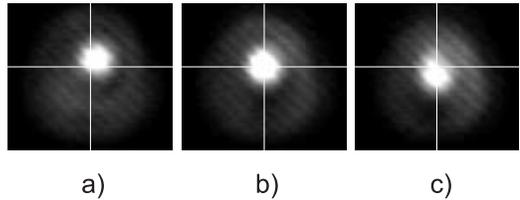}} \vspace{0.1cm}
\caption{ Motion of dark soliton. Recording time 3.5 $\mu$s (a), 5
$\mu$s (b), 6.5 $\mu$s (c) after start of illumination.}
\end{figure}
 The long formation time of these
solitons appears related to "critical slowing". We found that the
formation time for the solitons is reduced by a factor of 10 upon
increase of the illuminating intensity by only 10 $\%$. \\
\begin{figure}[htbf]
\epsfxsize=65mm \centerline{\epsfbox{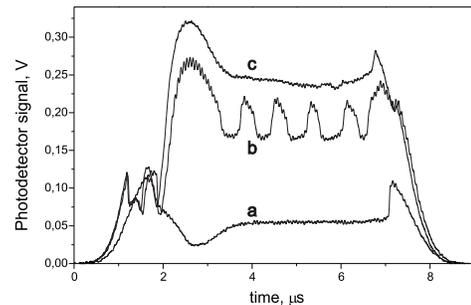}} 
\caption{ Reflected light measured at the center of the Gaussian
beam. Parameters are: $\delta \lambda$ = -0.8 nm, power 130 mW
(a), 245 mW (b) and 270 mW (c).}
\end{figure}
Fig. 6 gives the field dynamics for fixed detuning $\delta\lambda$
of the laser frequency from the resonator resonance frequency and
increasing light intensities. Fig. 6a) corresponds to the bright
soliton Fig. 1b) for small intensity. Fig. 6b) shows an unstable
(pulsing) dark soliton at medium intensity and 6c) gives the case
of the stable dark soliton as in Fig. 1c) for high intensity. Note
that the transient period between switching of the resonator (t =
1.3 $\mu$s for 6b), 6c)) and the onset of the dark soliton is here
reduced compared to Figs 3,4 due to the higher intensity (critical
slowing). The two minima of the reflected light before the onset
of the soliton in Fig. 4 a) which are related to spatial structure
(Fig. 4b)) also appear in Fig. 6b),c).\\
\begin{figure}[htbf]
\epsfxsize=70mm \centerline{\epsfbox{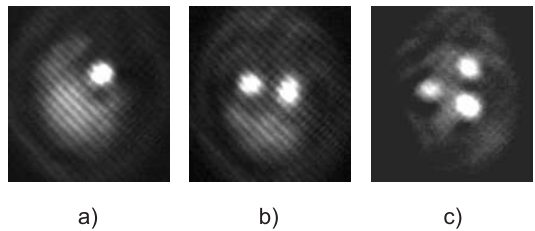}} \vspace{0.1cm}
\caption{Dark solitons observed in the switched area (intensity).}
\end{figure}
In the Figs 1-6 we have shown the cases where only one soliton
exists. In general, at higher intensities / at later times in the
illumination, several solitons can exist. Fig. 7 shows as an
illustration 1, 2, 3 solitons existing simultaneously. We have
observed up to 5 simultaneous solitons.\\

Concluding, we find at wavelengths corresponding to a wavelength
detuning of about one exciton linewidth above the exciton line
center and at high (negative) resonator detuning the existence of
bright and dark resonator solitons as predicted in \cite{tag:1}
and \cite{tag:2} respectively. The bright solitons appear stable,
while the dark solitons are observed to move in space and to
pulse, which makes them appear as less stable objects than the
bright solitons. \\

Acknowledgements \\ This work was supported by ESPRIT LTR project
PIANOS. \\

\end{document}